\documentclass[prl,aps,amssymb,showpacs,twocolumn]{revtex4}
\usepackage{epsfig}
\usepackage{amsmath}
\usepackage{amssymb}
\usepackage{amsthm}
\usepackage{amsfonts}
\usepackage{enumerate}
\usepackage{latexsym}

\newcommand{\beq}{\begin{equation}}
\newcommand{\eneq}{\end{equation}}
\def\avg#1{\langle#1\rangle}
\def\Re {\mbox{Re}}
\def\Im {\mbox{Im}}
\def\be{\begin{equation}}       \def\ee{\end{equation}}
\def\bea{\begin{eqnarray}}      \def\eea{\end{eqnarray}}

\def\PRB{Phys. Rev. B}

\def\PRL{Phys. Rev. Lett.}

\begin{document}

\title{Spin Current in  Spin-Orbit Coupling Systems  }
\author { Jiangping Hu}
\affiliation{Department of Astronomy and Physics, UCLA,
Los Angeles, California 90095}
\author{ Bogdan A. Bernevig}
\affiliation{ Department of Physics,
Stanford University, Stanford, California 94305 }
\author{Congjun Wu}
\affiliation{ Department of Physics,
Stanford University, Stanford, California 94305 }

\begin{abstract}
We present a simple and pedagogical derivation of the spin current
as the linear response to an external electric field for both
Rashba and Luttinger spin-orbital coupling Hamiltonians. Except
for the adiabatic approximation, our derivation is exact to the
linear order of the electric field for both models. The spin
current is a direct result of the difference in occupation levels
between different bands. Moreover, we show a general topological
spin current can be defined for a broad class of spin-orbit
coupling systems.

\end{abstract}
\maketitle

Spintronics aims to manipulate spins of particles. As such, an
essential step in the field is the generation of a reliable spin
current. The injection of spin polarized electron current from a
ferromangetic metal is not favorable because polarization is lost
at the interface due to the conductance mismatch \cite{hammar1999,
schmidt2000}. Injections from ferromagnetic semiconductors into
nonmagnetic semiconductors have been successfully developed in
recent several years \cite{ohno1999, fiederling1999, mattana2003}.
The theory of spin transport in the fore-mentioned cases depends
on the detailed mechanism of spin relaxation, where spin transport
generally is a dissipative process.

Recently, Murakami, Nagaosa and Zhang\cite{murakami2003} have
discovered a dissipationless and topological spin Hall current in
the hole doped semiconductors with strong spin orbit coupling.
These authors studied the effective Luttinger Hamiltonian
\cite{luttinger1956}
\begin{eqnarray}
 H_L= \frac{1}{2m}[(\gamma_1+\frac{5}{2}\gamma_2)
P^2-2\gamma_2( P\cdot  S)^2].
\end{eqnarray}
which describes conventional semiconductors such as Si, Ge, GaAs
and InSb. The geometrical structure of the effect is such that for
an electric field applied on the $z$ direction, a $y$-polarized
spin current will flow in the $x$ direction. The electric field
induced spin current can be summarized by the following formula
\begin{eqnarray}
J_j^i=\sigma_s\epsilon_{ijk}E_{k},
\end{eqnarray}
where $\epsilon_{ijk}$ is the antisymmetric tensor. Unlike the
ordinary Ohm's law, this equation has the remarkable property of
time reversal symmetry\cite{murakami2003,murakami2003A}. This
effect also has a deep topological
origin\cite{murakami2003,murakami2003A}, related to the
topological structure of the four dimensional quantum Hall
effect\cite{zhang2001}.

In the other independent work by Sinova et al\cite{sinova2003},
  the dissipationless, or the intrinsic
spin current is also discovered  to exist in the two dimensional
Rashba system, described by the Hamiltonian\cite{bychkov1984,
rashba}
\begin{eqnarray}
H_R=\frac{ P^2}{2m}+\gamma (P_xS_y-P_yS_x).
\end{eqnarray}
In this case, the spin current is polarized in the direction
perpendicular to the two dimensional plane and flowing in a planar
direction perpendicular to the direction of the charge current.
Surprisingly, the spin conductance in these systems turns out to
be independent of the spin orbit coupling and given by:
\begin{eqnarray}
\sigma_{s}=\frac{e}{8\pi}. \label{hallcurrent}
 \end{eqnarray}

In both of the above models, the intrinsic spin current is induced
by an external electric field. The authors of \cite{sinova2003,
murakami2003} have argued that the spin current is dissipationless
because the spin conductance is invariant under time-reversal
operations. However, the approaches taken by the authors of
\cite{sinova2003} and \cite{murakami2003} are markedly different.
In the first two-dimensional model \cite{sinova2003}, a
semi-classical approach is used to derive the spin current, while
in the second three-dimensional model \cite{murakami2003}, the
spin current is derived as a topological effect in momentum space.
It therefore appears that the physics of the spin current in these
two models is different. In fact, in the first model, there is no
known topological structure. However, from a purely theoretical
point of view, we should be able to derive the results and
understand the physics in one unified approach.

The aim of this article is to present a simple, pedagogical and
unified derivation of the spin current for both models. The
derivation of Murakami et al. \cite{murakami2003,murakami2003A}
emphasizes on the momentum space topology which requires some
advanced mathematical knowledge. The semi-classical derivations
presented in \cite{murakami2003,sinova2003,culcer2003} may not be
familiar to general readers. In view of the importance of the
effect, we feel that the general reader could benefit from a
simple derivation based on the standard time dependent
perturbation theory within single particle quantum mechanics. As
in \cite{murakami2003,sinova2003}, we work within the adiabatic
approximation. Since there is no interaction between particles,
the result derived here is exact as far as only the linear
response to the external electric field in the adiabatic limit is
concerned. Moreover, general conclusions about the spin current
can be manifestly drawn in this approach. Following the common
definition of the spin current, we show that the spin current is
always a direct result of the difference in occupation levels
between different bands in the models. Part of the spin current
can be interpreted as topological spin current. In fact, we
present an analysis and derivation of the topological spin current
for a broad class of spin-orbital coupling models.

In the following, we first calculate the spin current in the
Rashba and Luttinger models and then discuss the topological part
of the spin current. For the Luttinger Hamiltonian, we show that
the expectation value for the spin current in the heavy hole and
light hole states differs by exactly a minus sign. This result
leads to the conclusion that the contributions to the spin current
from the heavy hole and light hole bands should be exactly
opposite and differ by the Fermi velocities of two bands. The
total spin current gives a quantum correction to the semiclassical
result of Ref.~\cite{murakami2003}. The nature of the quantum
correction can be manifestly understood in our calculation, and
has also been discussed in Ref.~\cite{murakami2003A}. For the
Rashba Hamiltonian our approach gives the same spin current as the
Ref.~\cite{sinova2003} if the same definition of the spin current
is taken.

In our calculation, the spin current is defined by the velocity
times the spin, which is a rank two tensor. However, the velocity
operator in general does not commute with the spin operator in a
model with spin-orbit coupling. In order to define the spin
current tensor as a Hermitian operator, we have to symmetrize it:
\begin{eqnarray}\label{defspcurrent}
J_i^j= \frac{1}{2}(S_i\frac{\partial H}{\partial
P_j}+\frac{\partial H}{\partial P_j}S_i).
\end{eqnarray}

Let us consider a general spin-orbit coupling model described by
the many-body Hamiltonian $H( P, S)$. In the presence of a
constant external electric field, we choose the vector potential
$\vec A=-\vec E t$. The total Hamiltonian becomes time dependent,
 $H(t)=H(P-e\vec E t,  S)$. Let $ |G, t>$ be an instantaneous
ground state eigenstate of the time-dependent Hamiltonian,
\begin{eqnarray}
H(t) |G,t>=E_{G}(t)|G, t>.
\end{eqnarray}
The many-body ground state wavefunction $\Psi_G(t)$ of the
Hamiltonian satisfies the Schrodinger equation,
\begin{eqnarray}
i\frac{d}{dt}\Psi_G(t) =H(t)\Psi_G(t).
\end{eqnarray}
By first-order time-dependent perturbation theory, we have
\begin{widetext}
\begin{eqnarray}  \label{kubo}
&&|\Psi_{G}(t)\rangle=\exp\{ -i \int_0^t dt^\prime
E_G(t^\prime)\}  \Big \{ |G,t \rangle
+ i \sum_n \frac{|n,t\rangle \avg{n,t| \frac{\partial}{\partial t} |G,t}}
{E_n(t)-E_G (t)} (1-e^{i(E_n(t)- E_G(t))t}) \Big \},
\end{eqnarray}
where $|n,t\rangle$ are the instantaneous excited eigenstates.

The second term in the RHS of the expression above contains a fast
oscillation term which averages to zero, and which we neglect
below. For the non-interacting Fermi system, the above expression
can be simplified into summation over all instantaneous single
particle eigenstates. Then for an arbitrary operator $O$, the
difference of expectation values between  the perturbed and
unperturbed states is  given in  \cite{fradkin1991}
\begin{eqnarray} \label{average}
\avg{\Psi_G, t|\hat O|\Psi_G, t}=i
\sum_{\epsilon_{\lambda,P}<E_f<\epsilon_{\lambda^\prime, P}}
\frac{ \avg{\lambda,P(t)|\hat O|\lambda',P(t)}
\avg{\lambda',P(t)|\frac{\partial}{\partial t}|\lambda,P(t)}
+\avg{\lambda,P(t)|\frac{\partial}{\partial t}|\lambda',P(t)}
\avg{\lambda',P(t)| \hat O|\lambda,P(t)} }
{\epsilon_{\lambda',P(t)}-\epsilon_{\lambda,P(t)}},
\end{eqnarray}
\end{widetext}
where $| \lambda,P(t)\rangle$ is the instantaneous eigenstate with
polarization $\lambda$ of the single particle Hamiltonian.

 The entire calculation of the spin current that follows relies
on the above Kubo formula. However, when the Hamiltonian has
degenerate states, we use the following convention: if a set of
states, \{$|\lambda,P(t)>$\}, are degenerate in energy, we can
always choose a complete orthogonal basis  of states in the set,
\{$|\alpha,P(t)>$\}, such that, for any two new different
orthogonal states $|\alpha_1,P(t)>$ and $|\alpha_2,P(t)>$, we
have,
\begin{eqnarray}\label{orth}
\avg{\alpha_1,P(t)|\frac{\partial}{\partial t}|\alpha_2,P(t)}=0.
\end{eqnarray}
In this case, the summation index in the formula does not include
the degenerate states and therefore, the formula is well defined.

Let us now consider the particular case of the Rashba Hamiltonian
in an external electric field. The time dependent Rashba
Hamiltonian is given by
\begin{eqnarray}
H_R(t)= \frac{ P(t)^2}{2m}+\gamma (P_x(t)S_y-P_y(t)S_x).
\end{eqnarray}
For a given $P(t)$, the instantaneous eigenstates are given by
\begin{eqnarray}
|\lambda, P(t)>=U_R|\lambda >,
\end{eqnarray}
with $\epsilon(P(t))=\frac{P^2(t)}{2m}+\gamma\lambda |P(t)|$,
$U_R=e^{-i\phi(t) S_z}, \phi(t)=\tan^{-1}\frac{P_y(t)}{P_x(t)}$
the azimuthal angle, and $|\lambda\rangle$ is the eigenstate of
$S_y$ with $S_y|\lambda \rangle =\lambda|\lambda\rangle$.
Therefore, \bea \frac{\partial} {\partial t}|\lambda,P(t)>&=&
-i\frac{d\phi(t)}{dt}S_z|\lambda,P(t)>, \eea with
$\frac{d\phi(t)}{dt}= e \epsilon_{ij}E_i\frac{P_j(t)}{P^2(t)}$,
where $\epsilon_{ij}$ is rank-2 antisymmetric tensor. By applying
the Kubo formula, for any Hermitian operator $\hat O$, we obtain
\begin{eqnarray}
&&\avg{\lambda,P(t)|\hat O|\lambda,P(t)}=2e\epsilon_{ij}
E_i\frac{P_j(t)}{|P(t)|^3}\nonumber\\
&& \hspace{-1cm}
\times \sum_{\lambda'\neq \lambda} \avg{\lambda', P(t)|S_z|\lambda, P(t)}
\frac{Re(\avg{\lambda,P(t)|\hat O|\lambda',P(t)})}
{\gamma(\lambda'-\lambda)}.
\end{eqnarray}
For the spin one half particles, the above formula is simplified
to
\begin{eqnarray}
\avg{\pm \frac{1}{2},P(t)|O|\pm \frac{1}{2},P(t)}=\pm C_o
\epsilon_{ij} E_i\frac{P_j(t)}{P(t)^3}
\end{eqnarray}
where $C_o=e/ \gamma Re\avg{\frac{1}{2}|\hat O(t)|-\frac{1}{2}}$.
The spin current operator when the spin is polarized in the
perpendicular direction to the $xy$ plane is given by
\begin{eqnarray}
J^z_i=\frac{P_i}{m}S_z+\frac{\gamma}{2}\epsilon_{ij}(S_zS_j+S_jS_z).
\end{eqnarray}
 It is easy to show that $Re\avg{\frac{1}{2}|\hat
J_i(t)|-\frac{1}{2}}=\frac{P_i}{m}\avg{\frac{1}{2}|
S_z|-\frac{1}{2}}=\frac{P_i}{2m}$. Considering the whole fermi
surface, we can easily calculate the  total spin current. Let's
take the electric field is in $x$ direction and consider the
current $J_y$.
\begin{eqnarray}
J_y=\frac{e}{8\pi\gamma m}\Delta P_f
\end{eqnarray}
where the $\Delta P_f$ is the difference of the fermi velocity for
the two bands. In this model, $\Delta P_f =m\lambda$. This yields
the result of the Eq.\ref{hallcurrent}.

We now turn to the discussion of the effective Luttinger
Hamiltonian. In the presence of the external electric field, the
time dependent effective Luttinger Hamiltonian is
\begin{eqnarray}
H_L(t)=\frac{1}{2m}[(\gamma_1+\frac{5}{2}\gamma_2)( P(t))
^2+2\gamma_2(P(t)\cdot S)^2].
\end{eqnarray}
For a given $P(t)$, the Hamiltonian has four instantaneous
eigenstates,
\begin{eqnarray}
&&
H(t)|P(t),\lambda>=\epsilon_{\lambda}(P(t))|P(t),\lambda\rangle,
\nonumber \\
&& \frac{P(t)\cdot S}{|P(t)|}|P(t),\lambda>=\lambda|P(t),\lambda
\rangle.
\end{eqnarray}
where $\epsilon_{\lambda}(P(t))=\frac{P^2(t)}{2m}(\gamma_1+
(\frac{5}{2}-2\lambda^2)\gamma_2)$. For $\lambda=\pm \frac{3}{2}$
and $\lambda=\pm \frac{1}{2}$, they are referred to as the heavy
hole band and light hole band respectively. The eigenstates can be
explicitly written as
\begin{eqnarray}
|P(t),\lambda>= U_L|\lambda>, \ \
U_L=e^{-i\phi(t)S_z}e^{-i\theta(t)S_y}|\lambda>,
\end{eqnarray}
where $\tan(\phi(t))= P_y(t)/P_x(t)$,
$\cos(\theta(t))=P_z(t)/|P(t)|$ and
$S_z|\lambda>=\lambda|\lambda>$.

Since the eigenstates are degenerate, we have to choose an
orthogonal basis and satisfy the Eq.~\ref{orth} in order to use
the Kubo formula. Without loss of generality, we choose the
electric field in the $z$ direction. In this case, $\phi$ is
time-independent. Therefore,
\begin{eqnarray}
\avg{P(t),\lambda'|\frac{\partial}{\partial t}|P(t),\lambda}-i\frac{d\theta(t)}{dt}\avg{\lambda'|S_y|\lambda}
\end{eqnarray}
\noindent From this equation, we obtain that for the states with
the helicity equal to $\pm\frac{3}{2}$, the matrix element,
$\avg{P(t),-\frac{3}{2}|\partial_t|P(t),\frac{3}{2}}$  vanishes.
The only states for which we have to find an orthogonal base are
in the helicity $\pm \frac{1}{2}$. Let us define
\begin{eqnarray}
|P(t),+\rangle&=&\frac{1}{\sqrt{2}}|P(t),\frac{1}{2}\rangle
+i|P(t),-\frac{1}{2}\rangle, \nonumber \\
|P(t),-\rangle&=&\frac{1}{\sqrt{2}}(|P(t),\frac{1}{2}\rangle
-i|P(t),-\frac{1}{2}\rangle,
\end{eqnarray}
which satisfy $\avg{P(t),+|\partial_t|P(t),-} =0$.

For an arbitrary operator $\hat O$, let $\hat O(t)=U^+_L\hat O U_L$.
By applying the Kubo formula, we obtain the expectation value
for an arbitrary Hermitian operator,
\begin{eqnarray}
\avg{\Psi_{\frac{3}{2}}(t)|\hat O|\Psi_{\frac{3}{2}} (t)}=
\frac{\sqrt{3}m}{2\gamma_2P^2}\frac{d\theta(t)}{dt}
\Im(\avg{\frac{1}{2}|\hat O(t)|\frac{3}{2}})
\end{eqnarray}
and
\begin{eqnarray}
\avg{\Psi_{+}(t)|\hat O|\Psi_{+} (t)}&=&
\frac{\sqrt{3}m}{4\gamma_2P^2}\frac{d\theta(t)}{dt}
\Big\{\Im(\avg{\frac{3}{2}|\hat O (t) |\frac{1}{2}}+
\avg{-\frac{1}{2}|\hat O (t)|-\frac{3}{2}}) \nonumber \\
&&+ \Re(\avg{-\frac{1}{2}|\hat O (t) |\frac{3}{2}}
+\avg{\frac{1}{2}|\hat O (t) |-\frac{3}{2}})\Big \}
\end{eqnarray}
The spin current operator where the spin is polarized in $y$
direction and flows in $x$ direction is given by
\begin{eqnarray}
\hat
J_x^y=\frac{\gamma_1+\frac{5}{2}\gamma_2}{m}P_xS_y-\frac{\gamma_2}{2m}[
S_y((P\cdot S)S_x+S_x (P\cdot S))+h.c.].
\end{eqnarray}
The matrix element $\Im\avg{\frac{3}{2}|\hat J_x^y(t)|\frac{1}{2}}$ is
calculated to be
\begin{eqnarray}
\Im\avg{\frac{3}{2}|\hat J_x^y(t)|\frac{1}{2}}
=\frac{\sqrt{3}P}{2m}\sin\theta(\gamma_1 \cos^2\phi
+2\gamma_2 \sin^2\phi).
\end{eqnarray}
We thus obtain,
\begin{eqnarray}
&&<\Psi_{\frac{3}{2}}(t)|J_x^y|\Psi_{\frac{3}{2}}(t)>=-
<\Psi_{+}(t)|J_x^y|\Psi_{+}(t)>  = \nonumber \\ = & &
\frac{3e}{4\gamma_2P^4}(\gamma_1 P_x^2+2\gamma_2 P_y^2)E
\end{eqnarray}
where we used  $d\theta(t)/ dt= e E \sin (\theta(t))/|P(t)$.

For the states $|P(t),-\frac{3}{2}\rangle$ and $|P(t),-\rangle$,
the expectation values for spin current  are the same as
$|P(t),\frac{3}{2}\rangle$ and $|P(t),+\rangle$ respectively.
Therefore the total spin current by
including all the particles in the two bands is given by
\begin{eqnarray}
J_x^y= \frac{eE(\gamma_1+2\gamma_2)}{4\pi^2\gamma_2}\Delta P_f,
\end{eqnarray}
where $\Delta P_f$  is the fermi momentum difference between the
heavy and light hole bands.  Once again, we show that the spin
current comes from the occupation difference between two bands.
The above result has been independently obtained by Murakami,
Nagaosa and Zhang through a  slightly different derivation based
on Kubo formula too\cite{murakami2003A}.

This full quantum mechanical calculation gives a quantum
correction to the original semiclassical result
\cite{murakami2003}. In fact, the difference comes from the
definition of spin current operator. In \cite{murakami2003}, an
effective Hamiltonian was derived by introducing a monopole in
momentum space. The spin current is thought of as a topological
effect of the monopole. In the heavy hole states, the gauge
potential is abelian while in the light hole states, the gauge
potential is nonabelian. However, the field strength in both bands
is abelian. For each helicity states, the field strength is given
by
\begin{eqnarray}
F_{ij}=[D_i,D_j]=\lambda(\lambda^2-\frac{7}{2})\epsilon_{ijk}\frac{P_k}{P^3},
\end{eqnarray}
where the operator $D_j=P_{h(l)}(U^+_L \partial_{p_j}U_L)$
($P_{h(l)}$ is the projection onto heavy (light) hole bands). This
gauge field modifies the semiclassical equation of motions as
\bea\label{EOM} v_{i,\lambda}= {\dot
X}_{i,\lambda}=\frac{P_i}{m_\lambda}+ F_{ik} E_k.
\end{eqnarray}
In \cite{murakami2003}, the spin current is derived by replacing
the spin operator by its expectation value in the helicity states
compared to the definition in the Eq.\ref{defspcurrent}.  If we
use the same replacement, it is straightforward to show that the
spin current from  the perturbation theory is the same as from the
\cite{murakami2003}. Namely,  for a given helicity state
$|\lambda,P(t)>$, the expectation value of $<v_i>_{\lambda}$ in
the adiabatic approximation from the perturbation theory is given
by $ \frac{\partial P_j}{\partial t} <\lambda|F_{ij}|\lambda>$.

However, the above calculation does not underline the topological
nature. Several questions still remain. The first is how the
topological spin current can be separated from the general spin
current formula in the Eq.\ref{defspcurrent}. The second is that
since the calculation is performed on a specific model, it is not
clear whether the topological arguments can be applied to more
general cases such as realistic anisotropic, inversion-symmetry
breaking semiconductors. The recent work of Murakami Nagaosa and
Zhang\cite{murakami2003A} have answered some of these questions
from the Kubo formula for the isotropic Luttinger Hamiltonian.
Here we give an independent argument based on our formalism.

Let us review Eq.\ref{defspcurrent} and discuss a general case.
Let's assume a general unitary transformation $U$ which is a
function in the momentum space and diagonalizes a general
Hamiltonian $H$ . For any operator $O$, let $O(U)=U^+OU$.
Therefore, $H_0=H(U)$ is diagonal and
\begin{eqnarray}
J_i^j(U) =\frac{-i}{2}[S_i(U)[X_j(U),H_0]+[X_j(U),H_0]S_i(U)]
\label{rspincurrent}
\end{eqnarray}
Let's write $S_i(U)=S^p_i(U)+S^c_i(U)$, and
$X_j(U)=X_j^p(U)+X_j^c(U)$ where $O^p(U)$ keeps the elements of
$O(U)$ which are only between the degenerate eigenvalues of $H(U)$
for a given operator $O$. Namely  it is the projection onto the
degenerate bands. $O^c(U)$ is the leftover part. Now we can define
the total spin current operator into $J_i^j(U)=T_i^j(U)+
A_i^j(U)$, where the first part $T_i^j(U)$ is defined
\begin{eqnarray}
T_i^j(U)&=&\frac{-i}{2}([S^p_i(U)X_j(U)+X_jS^p_i(U),H_0]\nonumber \\
&+ &[S_i(U)X^p_j(U)+X^p_jS_i(U),H_0])
\end{eqnarray}
and
\begin{eqnarray}
 A_i^j(U)=\frac{-i}{2}[S^c_i(U)[X^c_j(U),H_0]+[X^c_j(U),H_0]S^c_i(U)].
\end{eqnarray}
where the  relations  $[S^p_i(U),H_0]=0$ and $[X_j^p(U),H_0]=0$
have been used in the above equations. It is clear that $A_i^j(U)$
is the band crossing contribution to the spin current. $T_i^j(U)$
can be considered as  the topological part of the spin current.
This statement is true for any models with arbitrary number of
bands and with arbitrary degeneracy in each bands caused by spin
orbit coupling. The proof is straightforward from the perturbation
theory.

 Without loss of generality, we assume that
$H_0$ is  the diagonal matrix, $H_0=\left (\begin{array}{cccc}
E_1I_{m_1} & 0  & 0 &...\\ 0 &E_2 I_{m_2} &0 &...\\ ...& ... &...
&...
\end{array} \right )$, where $m_1, m_2, ...$ are the number of degeneracies of each of the
bands. By a direct calculation from the Eq.\ref{defspcurrent} ,
the spin current  contribution from $T_i^j(U)$ is given by
\begin{eqnarray}
<T_i^j>= i Tr \{S^p_i(U)[X^p_j(U),X^p_k(U)]\}\frac{\partial
P_k(t)}{\partial t}.
\end{eqnarray}
The above equation is independent of $U$ for all of unitary
matrixes which  $U^+HU=H_0$.  Therefore, the symmetry group for
$U$ is $SU(m_1)\bigotimes SU(m_2)\bigotimes ...$. The above
formula is manifestly gauge invariant if we view the symmetry
group as a gauge group in momentum space as described in
\cite{murakami2003}.

From the above analysis, we see that the topological spin current
exists in  much broader spin-orbit coupling systems. However, it
requires the degeneracy of the bands, namely the non-abelian gauge
structure in momentum space.  The direct consequences from this
result is that for a realistic anisotropic Luttinger Hamiltonian,
the topological part of spin current will still exist, and that
for the Rashba Hamiltonian there is no topological part of the
spin current since there is only a $U(1)$ gauge.

 We would like to express our deep gratitude to S.C. Zhang for extensive
 discussions. JP
 would like to gratefully acknowledge the extreme useful discussion with
 YB. Kim and  ZQ. Wang during his stay at the Aspen Center of Physics.
 and he also would like to thank S. Murakami, L Balents,
  C. Nayak, L, Radzihovsky, C.M.
 Varma, Y. Bazaliy  and N. Sinitsyn for useful conversations.
 B.A.B would like to thank  R.B. Laughlin for stimulating
 discussions. CJ.W. and B.A.B are supported by the Stanford Graduate
 Fellowship program. JP is supported by the funds from the David
 Saxon chair at UCLA.

%The momentum-dependent gauge potential $\vec{A}_\lambda$ is

%identical to the one discussed by Zee [\cite{Zee}], as

%Up to the first order on $\vec E$,

%\begin{eqnarray}

%\frac{d\vec n(t)}{dt} =\frac{-e\vec E}{P(t)}-\vec

%P(t)\frac{d}{dt}\frac{1}{P(t)}.

%\end{eqnarray}

%Since $\vec{P}(t) \cdot \vec{A}_\lambda =0$, the effective

%hamiltonian becomes:

%\begin{eqnarray}

%H_{eff}(t)=\frac{\hat P(t)^2}{2m_\lambda} + {e \vec E} \frac{i

%\vec {A}_\lambda}{P(t)}

%\end{eqnarray}

%Upon the  change of notation $\frac{i \vec {A}_\lambda}{P(t)} \rightarrow

%\vec{A}_\lambda $, $\vec{A}_\lambda$ becomes identical to

%the one in reference \cite{zhang}.

\end{document}